\documentclass[prd,onecolumn,nofootinbib]{revtex4}
\usepackage{bm,amsmath,amssymb,graphicx}

\begin{document}

\title{Gravitational deflection of light in interior Schwarzschild metric}
\author{Nikodem Pop{\l}awski}
\altaffiliation{NPoplawski@newhaven.edu}

\affiliation{Department of Mathematics and Physics, University of New Haven, West Haven, CT, USA}

\begin{abstract}
We determine the angle of deflection of light by the gravitational field inside and outside a spherical body with a homogeneous mass density.
We show that the largest deflections, which can be measured by weak gravitational lensing, are in a region displaced from the center of mass.
This result can be extended to more general distributions of matter.
This displacement, observed in galaxies and colliding galaxy clusters, may be therefore explained without dark matter, within general relativity.
\end{abstract}
\maketitle

\noindent
{\bf Interior Schwarzschild solution}.\\
Gravitational deflection of light is one of the three classical tests of Einstein's general theory of relativity, beside precession of gravitational orbits and gravitational redshift \cite{GR,LL2}.
The classical formula for the deflection exists for the Schwarzschild metric, describing the gravitational field outside a centrally symmetric body \cite{Schw1}.
In this article, we derive the formula for gravitational deflection for the interior Schwarzschild metric, describing the gravitational field inside a spherical body with a homogeneous mass density \cite{Schw2,Lord}.

A centrally symmetric gravitational field is given by the metric in the spherical coordinates:
\begin{equation}
ds^2=e^{\nu(\tau,R)}c^2 d\tau^2-e^{\lambda(\tau,R)}dR^2-e^{\mu(\tau,R)}R^2(d\theta^2+\sin^2\theta\,d\phi^2),
\label{central}
\end{equation}
where $\nu$, $\lambda$ and $\mu$ are real functions of a time coordinate $\tau$ and a radial coordinate $R$.
Coordinate transformations $\tau\rightarrow\tilde{\tau}(\tau)$ and $R\rightarrow\tilde{R}(R)$ do not change the form of this metric.
The components of the Einstein tensor corresponding to (\ref{central}) that do not vanish identically are
\begin{eqnarray}
& & G_0^0=-e^{-\lambda}\Biggl(\mu''-\frac{2}{R^2}+\frac{3}{4}\biggl(\mu'+\frac{2}{R}\biggr)^2-\frac{1}{2}\biggl(\mu'+\frac{2}{R}\biggr)\lambda'\Biggr)+\frac{1}{2}e^{-\nu}\biggl(\dot{\lambda}\dot{\mu}+\frac{\dot{\mu}^2}{2}\biggr)+\frac{e^{-\mu}}{R^2}, \nonumber \\
& & G_1^1=-\frac{1}{2}e^{-\lambda}\Biggl(\frac{1}{2}\biggl(\mu'+\frac{2}{R}\biggr)^2+\biggl(\mu'+\frac{2}{R}\biggr)\nu'\Biggr)+e^{-\nu}\biggl(\ddot{\mu}-\frac{1}{2}\dot{\mu}\dot{\nu}+\frac{3}{4}\dot{\mu}^2\biggr)+\frac{e^{-\mu}}{R^2}, \nonumber \\
& & G_2^2=G_3^3=-\frac{1}{4}e^{-\lambda}\Biggl(2\nu''+\nu'^2+2\mu''-\frac{4}{R^2}+\biggl(\mu'+\frac{2}{R}\biggr)^2-\nu'\lambda' \nonumber \\
& & +\biggl(\mu'+\frac{2}{R}\biggr)(\nu'-\lambda')\Biggr)+\frac{1}{4}e^{-\nu}(2\ddot{\lambda}+\dot{\lambda}^2+2\ddot{\mu}+\dot{\mu}^2-\dot{\lambda}\dot{\nu}-\dot{\mu}\dot{\nu}+\dot{\lambda}\dot{\mu}), \nonumber \\
& & G_0^1=\frac{1}{2}e^{-\lambda}\Biggl(2\dot{\mu}'+\biggl(\mu'+\frac{2}{R}\biggr)(\dot{\mu}-\dot{\lambda})-\nu'\dot{\mu}\Biggr),
\label{Einstein}
\end{eqnarray}
where the dot denotes differentiation with respect to $\tau$ and the prime denotes differentiation with respect to $R$ \cite{LL2,Tol}.

In a comoving frame of reference, the spatial components of the four-velocity are zero.
In this frame, the Einstein equations are
\[
G_0^0=\kappa\epsilon,\quad G_1^1=G_2^2=G_3^3=-\kappa p,\quad G_0^1=0,
\]
where $\epsilon$ is the energy density of matter, $p$ is its pressure, and $\kappa=8\pi G/c^4$.
The conservation law $G^i_{k;i}=0$, where the semicolon denotes covariant differentiation, gives
\begin{equation}
\dot{\lambda}+2\dot{\mu}=-\frac{2\dot{\epsilon}}{\epsilon+p},\quad \nu'=-\frac{2p'}{\epsilon+p},
\label{conservation}
\end{equation}
where the constants of integration depend on the allowed transformations $\tau\rightarrow\tilde{\tau}(\tau)$ and $R\rightarrow\tilde{R}(R)$.

For a spherically symmetric body with radius $a$ in hydrostatic equilibrium, the metric does not depend on the time coordinate and a coordinate transformation $R\rightarrow\tilde{R}(R)=r$ of the radial coordinate can be applied to set $\mu(R)=0$.
The metric (\ref{central}) reduces to
\begin{equation}
ds^2=e^{\nu(r)}c^2 dt^2-e^{\lambda(r)}dr^2-r^2(d\theta^2+\sin^2\theta\,d\phi^2).
\label{metric}
\end{equation}
The Einstein equations for the components (\ref{Einstein}) reduce to
\begin{eqnarray}
& & \frac{1}{r^2}-e^{-\lambda}\biggl(\frac{1}{r^2}-\frac{\lambda'}{r}\biggr)=\kappa\epsilon, \nonumber \\
& & \frac{1}{r^2}-e^{-\lambda}\biggl(\frac{1}{r^2}+\frac{\nu'}{r}\biggr)=-\kappa p,
\label{field}
\end{eqnarray}
where $\epsilon$ and $p$ depend only on the radial coordinate $r$.
Integrating the first equation in (\ref{field}) gives
\begin{equation}
e^{\lambda}=\biggl(1-\frac{\kappa}{r}\int_0^r\epsilon r^2 dr\biggr)^{-1}=\biggl(1-\frac{r_g(r)}{r}\biggr)^{-1},
\label{lambda}
\end{equation}
where
\begin{equation}
r_g(r)=\kappa\int_0^r\epsilon(r)r^2 dr
\label{radius}
\end{equation}
is the gravitational radius of the sphere of radius $r$ centered at the origin.

The second equation in (\ref{conservation}) and the second equation in (\ref{field}) yield
\[
p'=-\frac{1}{2}(\epsilon+p)\Biggr(e^{\lambda}\biggr(\frac{1}{r}+\kappa pr\biggr)-\frac{1}{r}\Biggr)=-\frac{1}{2r^2}(\epsilon+p)e^{\lambda}\bigl(r_g(r)+\kappa pr^3\bigr),
\]
which, upon substituting (\ref{lambda}), gives the Tolman--Oppenheimer--Volkoff equation \cite{Tol}:
\begin{equation}
\frac{dp}{dr}=-\frac{1}{2r^2}\frac{(\epsilon+p)\bigl(r_g(r)+\kappa pr^3\bigr)}{1-r_g(r)/r}.
\label{TOV}
\end{equation}
In the nonrelativistic limit, it reduces to $dp/dr=-Gm(r)\rho(r)/r^2$, where $\rho=\epsilon/c^2$ is the mass density and
$m(r)=4\pi\int_0^r\rho(r)r^2 dr$ is the mass of the sphere of radius $r$ centered at the origin.
This relation is the radial component of the nonrelativistic hydrostatic equation $\mbox{{\bf grad}}\,p=\rho{\bf g}$, where ${\bf g}$ is the gravitational acceleration vector on the surface of this sphere.

If the energy density inside the sphere is homogeneous, $\epsilon=\mbox{const}$, then (\ref{lambda}) gives
\begin{equation}
e^{\lambda(r)}=\biggl(1-\frac{1}{3}\kappa\epsilon r^2\biggr)^{-1},
\end{equation}
which, with (\ref{TOV}), leads to
\begin{equation}
p'=-\frac{1}{2}(\epsilon+p)\biggl(1-\frac{1}{3}\kappa\epsilon r^2\biggr)^{-1}\biggl(\frac{\epsilon}{3}+p\biggr)\kappa r.
\end{equation}
Integrating this equation with the condition that the pressure vanishes at the boundary of the sphere, $p(a)=0$, gives
\[
p(r)=\epsilon\frac{(1-\kappa\epsilon r^2/3)^{1/2}-(1-\kappa\epsilon a^2/3)^{1/2}}{3(1-\kappa\epsilon a^2/3)^{1/2}-(1-\kappa\epsilon r^2/3)^{1/2}}.
\]
The pressure is a decreasing function of $r$.
The condition that the pressure $p(0)$ at the center be finite gives $a>(9/8)r_g$.
Physically realistic equations of state obey an inequality $p\le\epsilon/3$.
Therefore, $p(0)\le\epsilon/3$, which yields $a\ge(9/5)r_g$.

The second equation in (\ref{field}) is integrated to
\[
e^{\nu(r)}=e^{\nu(a)}\biggr(1+\frac{p(r)}{\epsilon}\biggr)^{-2}.
\]
The continuity of the metric at the boundary of the sphere, $r=a$, requires
\[
e^{\nu(a)}=e^{-\lambda(a)},
\]
which is satisfied by the Schwarzschild metric (exterior solution).
Therefore, the square of the line element for the interior Schwarzschild metric (\ref{metric}) for a constant energy density is given by
\[
ds^2=\frac{1}{4}\bigl(3(1-\kappa\epsilon a^2/3)^{1/2}-(1-\kappa\epsilon r^2/3)^{1/2}\bigr)^2 c^2 dt^2-\frac{dr^2}{1-\kappa\epsilon r^2/3}-r^2(d\theta^2+\sin^2\theta\,d\phi^2).
\]

Using the gravitational radius (\ref{radius}) for the entire body:
\[
r_g=r_g(a)=\frac{1}{3}\kappa\epsilon a^3=\frac{8\pi G}{3c^4}\rho c^2 a^3=\frac{2GM}{c^2},
\]
where $M=(4/3)\pi a^3\rho$ is the mass of the body, gives \cite{Schw2,Lord}
\begin{equation}
ds^2=\frac{1}{4}\bigl(3(1-r_g/a)^{1/2}-(1-r_g r^2/a^3)^{1/2}\bigr)^2 c^2 dt^2-\frac{dr^2}{1-r_g r^2/a^3}-r^2(d\theta^2+\sin^2\theta\,d\phi^2).
\label{interior}
\end{equation}
On the surface of the body, $r=a$, this metric gives
\[
ds^2=(1-r_g/a)c^2 dt^2-\frac{dr^2}{1-r_g/a}-r^2(d\theta^2+\sin^2\theta\,d\phi^2),
\]
which coincides with the Schwarzschild metric.\\

\noindent
{\bf Motion in a central field}.\\
The motion of a particle is given by the geodesic equation:
\[
\frac{d^2x^i}{ds^2}+\Gamma^{i}_{kl}\frac{dx^k}{ds}\frac{dx^l}{ds}=0,
\]
where $\Gamma^{i}_{kl}$ are the Christoffel symbols, forming the Levi-Civita connection \cite{GR,LL2}:
\[
\Gamma^{i}_{kl}=\frac{1}{2}g^{im}(g_{mk,l}+g_{ml,k}-g_{kl,m}).
\]
The nonzero components of the metric tensor $g_{ik}$ in (\ref{metric}) are: $g_{00}=e^\nu$, $g_{rr}=-e^\lambda$, $g_{\theta\theta}=-r^2$, $g_{\phi\phi}=-r^2\sin^2\theta$, and the comma denotes partial differentiation with respect to a coordinate.

The geodesic equation for $x^i=\theta$ is
\[
\frac{d^2\theta}{ds^2}+2\Gamma^{\theta}_{r\theta}\frac{dr}{ds}\frac{d\theta}{ds}+\Gamma^{\theta}_{\phi\phi}\biggl(\frac{d\phi}{ds}\biggr)^2=0,
\]
which gives
\[
\frac{d^2\theta}{ds^2}+\frac{2}{r}\frac{dr}{ds}\frac{d\theta}{ds}-\sin\theta\cos\theta\biggl(\frac{d\phi}{ds}\biggr)^2=0.
\]
If $\theta=\pi/2$ then $d\theta/ds=0$ and $d^2\theta/ds^2=0$, so this relation is satisfied.
The motion of the particle therefore takes place in a plane $\theta=\pi/2$ (any plane can be represented by this equation), which is a property of the motion in a central field.
Consequently, the motion can be described in the polar coordinates $r,\phi$.

The geodesic equation for $x^i=\phi$ is
\[
\frac{d^2\phi}{ds^2}+2\Gamma^{\phi}_{r\phi}\frac{dr}{ds}\frac{d\phi}{ds}=0,
\]
which gives
\[
\frac{d^2\phi}{ds^2}+\frac{2}{r}\frac{dr}{ds}\frac{d\phi}{ds}=\frac{1}{r^2}\frac{d}{ds}\biggl(r^2\frac{d\phi}{ds}\biggr)=0,\quad r^2\frac{d\phi}{ds}=-u_\phi=\mbox{const}=-l.
\]
The corresponding component $p_\phi$ of the four-momentum $p_i=mcu_i$ is related to the conserved angular momentum $M$ of the particle \cite{LL2}:
\begin{equation}
p_\phi=mcu_\phi=-mcl=-mcr^2\frac{d\phi}{ds}=-M,
\label{angularmomentum}
\end{equation}
where $m$ is the mass of the particle and $u^i=dx^i/ds$ is its four-velocity.
This conservation results from the central character of the field.

The geodesic equation for $x^i=x^0=ct$ is
\[
c^2\frac{d^2t}{ds^2}+2c\Gamma^{0}_{0r}\frac{dt}{ds}\frac{dr}{ds}=0,
\]
which gives
\[
c^2\frac{d^2t}{ds^2}+c\nu'\frac{dt}{ds}\frac{dr}{ds}=ce^{-\nu}\frac{d}{ds}\biggl(e^\nu\frac{dt}{ds}\biggr)=0,\quad ce^\nu\frac{dt}{ds}=u_0=\mbox{const}=h,
\]
where the prime denotes differentiation with respect to $r$.
The corresponding component $p_0$ of the four-momentum is proportional to the conserved energy $E$ of the particle \cite{LL2}:
\begin{equation}
p_0=mcu_0=mch=mc^2 e^\nu\frac{dt}{ds}=\frac{E}{c}.
\label{energy}
\end{equation}
This conservation results from the static character of the field.

Instead of writing the metric geodesic equation for $x^i=r$, we use the identity $u^i u_i=g_{ik}u^i u^k=1$, which yields
\[
c^2 e^\nu\biggl(\frac{dt}{ds}\biggr)^2-e^\lambda\biggl(\frac{dr}{ds}\biggr)^2-r^2\biggl(\frac{d\phi}{ds}\biggr)^2=1.
\]
Substituting (\ref{angularmomentum}) and (\ref{energy}) into this relation gives the radial equation of motion:
\begin{equation}
\biggl(\frac{E}{mc^2}\biggr)^2e^{-\nu}-\biggl(\frac{dr}{ds}\biggr)^2 e^\lambda-\biggl(\frac{M}{mc}\biggr)^2\frac{1}{r^2}=1,\quad e^{-\nu}h^2-e^\lambda\biggl(\frac{dr}{ds}\biggr)^2-\frac{l^2}{r^2}=1.
\label{radial}
\end{equation}
Using $dr/ds=(dr/d\phi)(d\phi/ds)$ in (\ref{radial}) gives
\[
e^{-\nu}h^2-e^\lambda\frac{l^2}{r^4}\biggl(\frac{dr}{d\phi}\biggr)^2-\frac{l^2}{r^2}=1,\quad h^2-e^{\nu+\lambda}\frac{l^2}{r^4}\biggl(\frac{dr}{d\phi}\biggr)^2-e^\nu\frac{l^2}{r^2}=e^\nu.
\]
Instead of $r$, it is more convenient to use its inverse $u=1/r$, leading to
\[
h^2-e^{\nu+\lambda}l^2\biggl(\frac{du}{d\phi}\biggr)^2-e^\nu\frac{l^2}{r^2}=e^\nu,\quad \frac{h^2}{l^2}-fg\biggl(\frac{du}{d\phi}\biggr)^2-\frac{f}{r^2}=\frac{f}{l^2},
\]
where $f(r)=e^\nu$ and $g(r)=e^\lambda$.

Differentiating this relation with respect to $\phi$, using $df/d\phi=(df/dr)(dr/du)(du/d\phi)$ and $dr/du=-1/u^2$, gives
\[
-2\frac{du}{d\phi}\frac{d^2 u}{d\phi^2}fg+\Bigl(\frac{du}{d\phi}\Bigr)^2(fg)'\frac{1}{u^2}\frac{du}{d\phi}+\Bigl(\frac{f}{r^2}\Bigr)'\frac{1}{u^2}\frac{du}{d\phi}=-\frac{1}{l^2}f'\frac{1}{u^2}\frac{du}{d\phi}.
\]
Assuming a noncircular path, dividing this relation by $(du/d\phi)$ gives the equation of path in the polar coordinates of a particle in a central field:
\begin{equation}
2fgu^2\frac{d^2 u}{d\phi^2}-\Bigl(\frac{du}{d\phi}\Bigr)^2(fg)'-\Bigl(\frac{f}{r^2}\Bigr)'=\frac{1}{l^2}f'.
\label{path}
\end{equation}

The propagation of light is described by an equation analogous to the geodesic equation:
\[
\frac{d^2x^i}{d\Lambda^2}+\Gamma^{i}_{kl}\frac{dx^k}{d\Lambda}\frac{dx^l}{d\Lambda}=0,
\]
where the interval $s$ is replaced with the affine parameter $\Lambda$ \cite{LL2}.
The wave four-vector $k^i=dx^i/d\Lambda$ satisfies $k^i k_i=0$.
Instead of $M$ and $E$, the conserved quantities are $r^2d\phi/d\Lambda=-k_\phi$ and the proper frequency $ce^\nu dt/d\Lambda=k_0$.
The entire derivation is analogous to that for a particle, except that the term on the right side of (\ref{radial}) is 0 instead of 1.
Consequently, the equation of path of light in the polar coordinates in a central field is given by (\ref{path}) without the term on the right side:
\begin{equation}
2fgu^2\frac{d^2 u}{d\phi^2}-\Bigl(\frac{du}{d\phi}\Bigr)^2(fg)'-\Bigl(\frac{f}{r^2}\Bigr)'=0.
\label{light}
\end{equation}

\noindent
{\bf Propagation of light for the interior solution}.\\
If the gravitational radius of the body is much smaller than its radius, $r_g\ll a$, then it is sufficient to determine the propagation of light in its interior up to linear terms in a small quantity $r_g/a$.
Using $(1+x)^n\approx 1+nx$ for $|x|\ll1$, the interior Schwarzschild metric (\ref{interior}) gives
\[
f=\frac{1}{4}\bigl(3(1-r_g/a)^{1/2}-(1-r_g r^2/a^3)^{1/2}\bigr)^2\approx 1-\frac{3r_g}{2a}+\frac{r_g r^2}{2a^3},\quad \frac{f}{r^2}\approx\frac{1}{r^2}-\frac{3r_g}{2ar^2}+\frac{r_g}{2a^3},
\]
and
\[
g=(1-r_g r^2/a^3)^{-1}\approx 1+\frac{r_g r^2}{a^3},
\]
which yields
\[
\Bigl(\frac{f}{r^2}\Bigr)'\approx -\frac{2}{r^3}+\frac{3r_g}{ar^3},\quad fg\approx 1-\frac{3r_g}{2a}+\frac{3r_g r^2}{2a^3},\quad 2fgu^2\approx\frac{2}{r^2}-\frac{3r_g}{ar^2}+\frac{3r_g}{a^3},\quad (fg)'\approx\frac{3r_g r}{a^3}.
\]
Consequently, the equation of path (\ref{light}) gives
\[
\Bigl(\frac{2}{r^2}-\frac{3r_g}{ar^2}+\frac{3r_g}{a^3}\Bigr)\frac{d^2 u}{d\phi^2}-\Bigl(-\frac{2}{r^3}+\frac{3r_g}{ar^3}\Bigr)-\frac{3r_g r}{a^3}\Bigl(\frac{du}{d\phi}\Bigr)^2=0.
\]
Dividing this relation by $2u^2$ gives
\begin{equation}
\Bigl(1-\frac{3r_g}{2a}+\frac{3r_g}{2a^3 u^2}\Bigr)\frac{d^2 u}{d\phi^2}+\Bigl(u-\frac{3r_g}{2a}u\Bigr)-\frac{3r_g}{2a^3 u^3}\Bigl(\frac{du}{d\phi}\Bigr)^2=0.
\label{linear}
\end{equation}
For the exterior Schwarzschild metric, $f=1/g=1-r_g/r$, and the analogous relation is obtained without linearization in $r_g$ \cite{LL2}:
\begin{equation}
\frac{d^2 u}{d\phi^2}+u=\frac{3}{2}r_g u^2.
\label{exterior}
\end{equation}

\begin{figure}[th]
\centering
\includegraphics[width=2.5in]{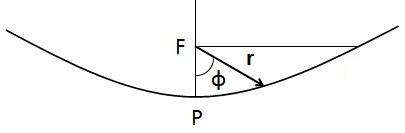}
\caption{Deflection of light.}
\label{deflection}
\end{figure}

In the absence of the field ($r_g=0)$, (\ref{linear}) reduces to $d^2 u/d\phi^2+u=0$, whose solution is
\begin{equation}
u_0=\frac{\cos\phi}{b},
\label{line}
\end{equation}
where $b$ (for the interior solution $b\le a$) is the impact parameter of a ray of light with respect to the origin at the center of the body.
Such a solution represents a straight line, passing by the origin at a distance $b$.
In the presence of the field, the path deviates from a straight line, as shown in Figure \ref{deflection}, in which $|FP|\approx b$, $F$ is the center of the body, and $P$ is the point of the closest approach of the ray passing the body.\footnote{
We assume that light does not interact with the matter in the body and its trajectory deflects only because of the curvature of spacetime.
}

We seek a solution of (\ref{linear}) in a form $u=u_0+u_1$, where $u_1\ll u_0$.
Substituting it into (\ref{linear}) and taking $u=u_0$ in all terms with $r_g$ ($u_1$ in terms with $r_g$ is of higher order of smallness) gives
\[
\Bigl(-\frac{3r_g}{2a}+\frac{3r_g}{2a^3 u^2_0}\Bigr)\frac{d^2 u_0}{d\phi^2}+\frac{d^2 u_1}{d\phi^2}+u_1-\frac{3r_g}{2a}u_0-\frac{3r_g}{2a^3 u^3_0}\Bigl(\frac{du_0}{d\phi}\Bigr)^2=0.
\]
Substituting (\ref{line}) into this relation yields
\begin{equation}
\frac{d^2 u_1}{d\phi^2}+u_1=\frac{3r_g b}{2a^3\cos^3\phi}.
\label{correction}
\end{equation}
A particular solution of this inhomogeneous linear equation can be determined by introducing a complex function $z=du_1/d\phi+iu_1$, giving $dz/d\phi-iz=3r_g b/2a^3\cos^3\phi$, whose solution is
\[
z=e^{i\phi}\int\frac{3r_g b}{2a^3\cos^3\phi}e^{-i\phi}d\phi=\frac{r_g b}{a^3}\frac{3ie^{i\phi}}{(e^{2i\phi}+1)^2},\quad u_1=\mbox{im}\,z=\frac{3r_g b}{4a^3\cos\phi}.
\]
Consequently, the first-order solution of (\ref{linear}) is:
\begin{equation}
u=\frac{\cos\phi}{b}+\frac{3r_g b}{4a^3\cos\phi}.
\label{solution}
\end{equation}
This equation determines the trajectory of light in the metric (\ref{interior}) in the polar coordinates as a function of the impact parameter $b$ and the radius of the body $a$.
For the exterior Schwarzschild metric, the analogous relation is \cite{LL2}
\[
u=\frac{\cos\phi}{b}+\frac{r_g}{2b^2}(2-\cos^2\phi).
\]

\begin{figure}[th]
\centering
\includegraphics[width=1.6in]{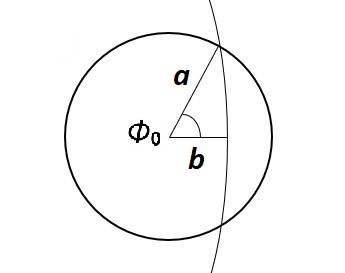}
\caption{Propagation of light in a spherical body.}
\label{inside}
\end{figure}

\noindent
{\bf Angle of deflection at the surface of a body}.\\
The polar angle $\phi_0$ at which a ray of light intersects the surface of a spherical body, as shown in Figure \ref{inside}, is determined from the equation of path (\ref{solution}) for $u=1/a$, giving a quadratic equation for $\cos\phi_0$:
\[
\cos^2\phi_0-\frac{b}{a}\cos\phi_0+\frac{3r_g b^2}{4a^3}=0.
\]
Its physical solution has a positive sign in front of the square root:
\begin{equation}
\cos\phi_0=\frac{b}{2a}\Bigl(1+\sqrt{1-\frac{3r_g}{a}}\Bigr)\approx\frac{b}{a}\Bigl(1-\frac{3r_g}{4a}\Bigr),\quad\sin\phi_0\approx\sqrt{1-\frac{b^2}{a^2}+\frac{3r_g b^2}{2a^3}}\approx\sqrt{1-\frac{b^2}{a^2}}+\frac{3r_g b^2/4a^3}{\sqrt{1-b^2/a^2}}.
\label{angle}
\end{equation}
In the absence of the field in the body, $\cos\phi_0=b/a$, corresponding to a triangle formed by the segments $a$ and $b$ with the vertical path of the ray.
The gravitational field deflects the ray, increasing $\phi_0$ and thus decreasing $\cos\phi_0$.
If a ray passes through the center of the body, then the impact parameter $b=0$, giving $\cos\phi_0=0$ and thus $\phi_0=\pi/2$.
Therefore, a ray passing through the center is not deflected by the gravitational field, which also follows from the central symmetry of the body.\footnote{
The approximation for the square root in $\sin\phi_0$ in (\ref{angle}) is valid if $a-b\gg r_g$.
Actually, if the distance of the closest approach is equal to the radius of the body, $b=a$, then $\phi_0$ should be zero.
However, the value of $\cos\phi_0$ in (\ref{angle}) show that $\phi_0$ in this case is a small positive quantity.
It is because the actual impact parameter is not $b$, but the distance between the center of the body and the asymptote reached by the ray at infinity.
The difference between the impact parameter and $b$ is of first order in $r_g$, but its effect on the angle of deflection is of higher order in $r_g$ and thus can be neglected.
}

\begin{figure}[th]
\centering
\includegraphics[width=1.4in]{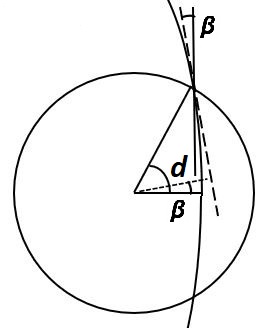}
\caption{Angle of deflection at the surface of a body.}
\label{outside}
\end{figure}

In the absence of the field outside the body, the path of a ray there would be a straight line.
Its slope in the polar coordinates can be determined using $dy/dx=[d(r\sin\phi)/d\phi]/[d(r\cos\phi)/d\phi]=(r'\sin\phi+r\cos\phi)/(r'\cos\phi-r\sin\phi)=(u'\sin\phi-u\cos\phi)/(u'\cos\phi+u\sin\phi)$ at $\phi=\phi_0$, where the prime denotes differentiation with respect to $\phi$.
Substituting (\ref{solution}) and
\begin{equation}
u'=\frac{du}{d\phi}=-\frac{\sin\phi}{b}+\frac{3r_g b}{4a^3}\frac{\sin\phi}{\cos^2\phi}
\label{derivative}
\end{equation}
gives
\[
\frac{dy}{dx}=\frac{1/b+(3r_g b/4a^3)(1-\tan^2\phi_0)}{-(3r_g b/2a^3)\tan\phi_0}.
\]
Putting here the values (\ref{angle}) gives the slope of the path of a ray at the surface of the body, up to terms of the first order in $r_g$:
\[
\frac{dy}{dx}=\tan(\pi/2+\beta)=-\cot\beta=\frac{a/b-b/a-2a^2/3r_g b}{\sqrt{1-b^2/a^2+3r_g b^2/2a^3}}<0,
\]
where $\beta$ is the angle between the tangent to the path at the surface of the body (long-dashed line) and the vertical, as shown in Figure \ref{outside}.
This angle, using $r_g\ll a$, is equal to
\begin{equation}
\beta=\arctan\Bigl(\frac{3r_g b}{2a^2}\sqrt{1-\frac{b^2}{a^2}}\Bigr)\approx \frac{3r_g b}{2a^2}\sqrt{1-\frac{b^2}{a^2}}.
\label{surface}
\end{equation}

A ray passing through the center, $b=0$ (corresponding to $\phi_0=\pi/2$), is not deflected at the surface: $\beta=0$.
This result follows from the central symmetry of the body.
A ray passing the body on its surface, $b=a$, is also not deflected at the surface because $\phi_0=0$: the interior metric does not contribute to the deflection.
The largest deflection at the surface occurs for $b=a/\sqrt{2}$ (corresponding to $\phi_0=\pi/4$): $\beta=3r_g/4a$.
The contribution to the total angle of deflection arising from the interior gravitational field is the angle between the tangents at the two intersections of a ray with the surface, which is equal to $2\beta$.\\

\noindent
{\bf Total angle of deflection}.\\
The deflection of light propagating outside the body can be determined from the first radial equation (\ref{radial}) for the exterior Schwarzschild metric with 0 instead of 1 on the right side, and taking the limit $m\to 0$.
This equation is equivalent to
\[
E^2-(Mc)^2 u^2(1-r_g u)-(Mc)^2\Bigl(\frac{du}{d\phi}\Bigr)^2=0.
\]
The impact parameter $d$ for the exterior trajectory starting at the surface of a body is represented by the dashed line in Figure \ref{outside}.
It is given by $M=dp=dE/c$, where $p$ is the momentum of a particle of light:
\begin{equation}
d=\frac{Mc}{E},\quad 1-d^2 u^2(1-r_g u)-d^2\Bigl(\frac{du}{d\phi}\Bigr)^2=0,\quad \frac{du}{d\phi}=-\sqrt{1/d^2-u^2(1-r_g u)},
\label{impact}
\end{equation}
where the minus sign of the square root is taken because $u$ decreases as $\phi$ increases (for the upper part of a trajectory).
Consequently, the change of the polar angle as the light propagates from $u=1/a$ at the surface to $u=0$ at infinity is given by
\[
\Delta\phi=\int_0^{1/a}\frac{du}{\sqrt{1/d^2-u^2(1-r_g u)}}.
\]
The corresponding angle of deflection outside the body is given by the difference between the path with $r_g$ and the path without $r_g$ (straight line):
\[
\delta\phi=\int_0^{1/a}\frac{du}{\sqrt{1/d^2-u^2}}-\int_0^{1/a}\frac{du}{\sqrt{1/d^2-u^2(1-r_g u)}}.
\]
Because this angle is on the order of magnitude of $r_g$, $d$ can be approximated by $b$.\footnote{
Up to terms of first order in $r_g$, $d$ can be calculated by substituting the derivative (\ref{derivative}) for $\phi=\phi_0$ (\ref{angle}) and $u=1/a$ into (\ref{impact}), yielding
$d=b(1+3r_g/4a-r_g b^2/a^3)$.
It can also be calculated, using $d=a\cos(\phi_0-\beta)\approx a(\cos\phi_0+\beta\sin\phi_0)$ in Figure \ref{outside}, yielding
$d=b(1+3r_g/4a-3r_g b^2/2a^3)$.
The difference between these two values arises from the approximate character of (\ref{angle}) and is of order $r_g$.
Therefore, replacing $d$ by $b$ in $\delta\phi$ introduces additional terms of higher order in $r_g$, which can be omitted in the linear approximation in $r_g$.
}

Finally, the total angle deflection for the entire trajectory inside and outside the body is equal to the sum of $2\beta$ (\ref{surface}) and $2\delta\phi$, where the factor 2 corresponds to a ray coming from infinity to the body (lower part of a trajectory) and then to infinity (upper part of a trajectory):
\begin{equation}
\delta\phi=\frac{3r_g b}{a^2}\sqrt{1-\frac{b^2}{a^2}}+2b\int_0^{1/a}\frac{du}{\sqrt{1-b^2 u^2}}-2b\int_0^{1/a}\frac{du}{\sqrt{1-b^2 u^2(1-r_g u)}}.
\label{final}
\end{equation}
This formula is valid for $0\le b \le a$.
For a ray passing through the center, $b=0$ gives $\delta\phi=0$.
Such a ray is not deflected, which follows from the central symmetry of the body.
In the limit $a\to 0$, $b\to 0$ and $r_g\sim a^3\to 0$, giving $\delta\phi\to 0$.
For $b>a$, the angle of deflection is given by the Einstein formula \cite{GR,LL2}:
\[
\delta\phi=2b\int_0^{1/b}\frac{du}{\sqrt{1-b^2 u^2}}-2b\int_0^{1/b}\frac{du}{\sqrt{1-b^2 u^2(1-r_g u)}}\approx\frac{2r_g}{b}.
\]
For a ray passing the body on its surface, $b=a$ gives $\delta\phi\approx 2r_g/a$, which also follows from (\ref{final}).\\

\noindent
{\bf Displacement between the centers of mass and gravity}.\\
According to the formula (\ref{final}), the angle of deflection $\delta\phi(b)$ inside a spherical body with a constant density of matter is an increasing function of the impact parameter $b$.
Outside the body, it is a decreasing function of $b$. 
At the center of the body, which is the center of mass of the body, the angle of deflection is equal to zero.
The region with largest deflections is near its surface, away from the center.
Consequently, the center of gravity, measured by weak gravitational lensing, is displaced from the center of visible mass.
In a galaxy, this observed displacement may be therefore explained without dark matter in a galactic halo, within the general theory of relativity.
A similar displacement can be obtained for more general distributions of mass, which are not homogeneous and spherical.
For an ellipsoidal body, the largest deflections will be in the regions along the longest axis of the ellipsoid and near its surface, away from the center of mass.
Accordingly, in a colliding cluster of two galaxies, the largest deflections will be along the axis connecting the galaxies, farther away from the middle point between the galaxies than the galaxy centers of mass.
This observed displacement may be therefore explained without dark matter in colliding galaxies, within general relativity or its extension with spin angular momentum: Einstein--Cartan theory gravity \cite{EC}, which may remove gravitational singularities in black holes and in the Universe \cite{reg}.

I am grateful to Francisco Guedes and my Parents, Bo\.{z}enna Pop{\l}awska and Janusz Pop{\l}awski, for their support.

\end{document}